# Impact-Parameter Description of High-Energy Deuteron-Nucleus Collisions

V. Franco[a*2] and R.J. Glauber[b1]

[a]Physics Department, Brooklyn College of the City University of New York, Brooklyn, New York 11210
[b]Lyman Laboratory of Physics, Harvard University, Cambridge, Massachusetts 02138

A theoretical analysis using an impact-parameter description of the collisions of deuterons with nuclei is carried out in the high-energy diffraction approximation. It is used to obtain the intensities and integrated cross sections for elastic scattering, for the emergence of the two incident nucleons from the collision whether they appear as an elastically scattered deuteron or as two unbound nucleons, and for the diffraction-induced dissociation of the deuteron into a free neutron and a free proton, as well as the total cross section. The cross section for collisions in which one or both of the nucleons is absorbed is derived in terms of the sum of the neutron-nucleus and proton-nucleus effective phase shifts. Expressions for the cross section for processes in which the proton (or neutron) is absorbed whether the neutron (or proton) is absorbed or not, and for the cross section for processes in which the neutron (or proton) is absorbed and the proton (or neutron) remains free are derived. A reduced form of a two-particle density matrix is introduced to directly derive expressions for the cross section for two-particle absorption in which both the proton and neutron are absorbed and for the cross section for stripping processes in which the proton (or neutron) is absorbed and the neutron (or proton) emerges as a free particle. The expression for the cross section for the breakup of the deuteron and the resulting emergence of one or two free nucleons is also derived. The mechanism by which the diffraction dissociation of the deuteron is induced is understood in an approximate semi-quantitative basis in classical terms (primarily the radial derivative of the radial impulse), allowing an estimate of where in the nuclear potential (beyond the "radius", near the "surface") the dissociation process tends to predominantly occur.

*Corresponding author.
E-mail address: vfranco@brooklyn.cuny.edu
[2]Postal address (until College lockdown is lifted): 960 East 18th Street, Brooklyn, NY 11230

[1]Deceased

## 1.  Introduction

Collisions between particles and deuterons have been studied for many decades.  Glauber [1] formulated a theory for high-energy collisions, a high-energy approximation.  Franco and Glauber [2,3] presented a detailed application of this approximation to hadron-deuteron collisions.  As was indicated in [1], the theory is also applicable to collisions between an incident deuteron and a nucleus.  A detailed study of scattering, diffraction dissociation, and stripping, as well as of numerous deuteron-nucleus absorptive processes was made by Franco [2].  Applications were made to scattering and diffraction dissociation by Zamick [4] and to dissociation and stripping by Fäldt [5]. Subsequently numerous such applications to deuteron-nucleus collisions have been made.

In the present work we derive expressions, within the framework of the high-energy approximation, for the cross sections for various processes that may occur in deuteron-nucleus collisions. Many of the results were first given by Franco [2]. We will derive *directly*, without resorting to subtractions of other deuteron-nucleus cross sections, the cross section for absorption processes in which some incoherent collision takes place involving *both* the neutron *and* the proton by introducing a two-particle density matrix and a reduced density operator for the incident proton (or neutron) of the deuteron.

Among the various deuteron-nucleus scattering processes is diffraction dissociation, first noticed and described by Glauber [6]. We show that it can be understood on a semi-quantitative basis in semi-classical terms, given some approximations. In such a description we find that it is mainly the radial derivative of the radial impulse that governs the dissociation process at small scattering angles of the center of mass. We illustrate this result with an effective potential, and indicate that the diffraction dissociation process occurs predominantly in the surface region, but beyond the "radius" of the nucleus.

The analysis is begun in Sec. 2 by presenting an impact-parameter approach to collisions of high-energy deuterons with nuclei in which the various scattering (non-absorptive) cross sections are obtained. In Sec. 3 a semi-classical description of diffraction dissociation with some approximations is used to indicate the region of the nuclear potential in which the process tends to predominantly occur. In Sec.4 the effects of absorption and the various absorptive cross sections are obtained.

## 2. Impact parameter approach for scattering of deuterons

In this section we shall describe scattering of high-energy deuterons by nuclei. We shall use a formalism which is asymptotically correct for collisions at relatively small momentum transfers, which are the predominant varieties occurring at high energies. This formalism, the high-energy approximation, or a generalized form of the eikonal approximation, has been studied extensively [7] and applied to a wide variety of problems [8].

We begin by presenting an approximate form for the amplitude for high-energy elastic scattering of particles by arbitrary nuclei. We then utilize this expression to secure amplitudes for the scattering of deuterons by stationary target nuclei. Expressions for intensities and integrated cross sections for deuteron-nucleus scattering processes, such as elastic scattering and the diffraction-induced dissociation of the deuteron into a free neutron and a free proton, as well as for the deuteron-nucleus total cross section, are derived.

In describing collisions between deuterons with momentum $\hbar k$ and stationary nuclei, it is convenient to begin by considering collisions between nucleons with momentum $\frac{1}{2}\hbar k$ and stationary nuclei. It has been shown [1] that in the high-energy approximation the optical model provides a simple representation for the elastic scattering amplitude for nucleon-nucleus collisions in terms of an effective complex phase shift $\chi(\mathbf{b})$, where $\mathbf{b}$ is the impact parameter vector for the collision and lies in the plane perpendicular to the velocity of the incident beam. For a nucleon momentum $\frac{1}{2}\hbar k$ this representation for the nucleon-nucleus elastic scattering amplitude $f(\mathbf{q}, \frac{1}{2}k)$ is given by [1]

$$f(\mathbf{q}, \tfrac{1}{2}k) = \frac{ik}{4\pi} \int e^{i\mathbf{q}\cdot\mathbf{b}}[1 - e^{i\chi(\mathbf{b})}]\mathrm{d}^{(2)}\mathbf{b} , \tag{1}$$

where $\hbar\mathbf{q}$ is the momentum transferred to the target nucleus and the integration is taken over the plane of impact parameter vectors **b**. The phase shift function $\chi(\mathbf{b})$ depends upon the interaction between the incident nucleon and the target nucleus and upon the initial state of the nucleus.

To describe collisions between a deuteron and a nucleus we let $\mathbf{r}_n$ and $\mathbf{r}_p$ denote the position vectors of the neutron and proton so that $\mathbf{r} = \mathbf{r}_p - \mathbf{r}_n$ is the internal coordinate of the deuteron, and we let **b** now denote the projection of the deuteron center of mass position vector on the plane perpendicular to the velocity of the incident beam. The magnitude of **b** corresponds to the classical impact parameter of the collision. Let us imagine the internal coordinate of the deuteron to have the fixed value **r** during the course of the collision. Then the incident deuteron wave function, upon passing through the nucleus, will accumulate a total phase shift which depends upon the coordinate **r** as well as on the impact parameter vector **b**. If we write this total phase shift function for interactions of the deuteron with the nucleus as $\chi_{\text{tot}}(\mathbf{b}, \mathbf{r})$ then, by analogy with nucleon-nucleus collisions, the elastic scattering amplitude for deuteron-nucleus collisions *would* be given by

$$F(\mathbf{q}, k, \mathbf{r}) = \frac{ik}{2\pi} \int e^{i\mathbf{q}\cdot\mathbf{b}}[1 - e^{i\chi_{\text{tot}}(\mathbf{b},\mathbf{r})}]\mathrm{d}^{(2)}\mathbf{b} \tag{2}$$

where $\hbar k$ is the momentum of the incident deuteron and $\hbar\mathbf{q}$ is the momentum transferred to the nucleus. We might remark that this expression is identical in form to that for particle-deuteron collisions. If the target nucleus is represented schematically by a complex potential well, the major difference between a deuteron-nucleus collision and the collision of an incident particle with the deuteron lies in the frame of reference used [1].

The internal coordinate **r** of the deuteron is, of course, not actually fixed during the collision. We may take account of that fact by noting that $1 - \exp[i\chi_{\text{tot}}(\mathbf{b}, \mathbf{r})]$ may be regarded as an operator which induces changes in the internal state of the deuteron through its dependence on **r**, much as it changes the momentum state of the incident deuteron through its dependence on the coordinate **b**. The two-particle scattering amplitudes which we observe in deuteron-nucleus collisions are the matrix elements of $F(\mathbf{q}, k, \mathbf{r})$ between the initial and final internal states of the deuteron. If we represent these states by $|i\rangle$ and by $|f\rangle$, respectively, we may write the corresponding scattering amplitude as

$$F_{fi}(\mathbf{q}, k) = \langle f| F(\mathbf{q}, k, \mathbf{r}) | i \rangle \tag{3}$$

$$= \int \phi_f^*(\mathbf{r}) F(\mathbf{q}, k, \mathbf{r}) \phi_i(\mathbf{r}) \, \mathrm{d}\mathbf{r} \tag{4}$$

where, in the latter form, $\phi_i(\mathbf{r})$ and $\phi_f(\mathbf{r})$ are, respectively, the initial and final internal wave functions for the deuteron. The initial state will ordinarily be the deuteron ground state, but the final state $|f\rangle$ may be any state of the two-nucleon system. Processes in which the neutron and proton emerge from the collision in an unbound state correspond to cases in which $|f\rangle$ is an excited state of the two-nucleon system. Elastic scattering is the process which is described by choosing $|f\rangle$ to be the ground state.

The differential cross section $(d\sigma/d\Omega)_{fi}$ for a collision in which a deuteron of momentum $\hbar k$ transfers momentum $\hbar \mathbf{q}$ to the nucleus and is left in a final state $|f\rangle$ is given by

$$(d\sigma/d\Omega)_{fi} = |F_{fi}(\mathbf{q},k)|^2, \tag{5}$$

where we have implicitly neglected the velocity change of the deuteron center of mass. In particular, the angular distribution for elastic scattering $(d\sigma/d\Omega)_{el}$ is given by

$$(d\sigma/d\Omega)_{el} = (d\sigma/d\Omega)_{ii} \tag{6}$$

$$= |F_{ii}(\mathbf{q},k)|^2 \tag{7}$$

which may be expressed in terms of the total phase shift $\chi_{tot}(\mathbf{b}, \mathbf{r})$ by using Eq. (2) to write

$$\left(\frac{d\sigma}{d\Omega}\right)_{el} = \left(\frac{k}{2\pi}\right)^2 \int e^{i\mathbf{q}\cdot(\mathbf{b}-\mathbf{b}')} \langle 1 - e^{i\chi_{tot}(\mathbf{b},\mathbf{r})}\rangle\langle 1 - e^{-i\chi^*_{tot}(\mathbf{b}',\mathbf{r})}\rangle \, d^{(2)}\mathbf{b} \, d^{(2)}\mathbf{b}' \tag{8}$$

where the brackets $\langle \ \rangle$ denote expectation values taken in the deuteron ground state.

The summed scattered intensity for the emergence of the two incident nucleons from the collision, $(d\sigma/d\Omega)_2$, whether they appear as an elastically scattered deuteron or as two unbound nucleons, is obtained by summing the squared modulus of the amplitude $F_{fi}(\mathbf{q}, k)$ over a complete set of final internal deuteron states $|f\rangle$. We then have the intensity for the sum of the elastic scattering and diffraction-induced splitting or dissociation of the deuteron, which we may write as

$$(d\sigma/d\Omega)_2 = \Sigma_f |F_{fi}(\mathbf{q},k)|^2. \tag{9}$$

In the high-energy approximation we may neglect the energy differences of the various final states of the deuteron since they are nearly equal as long as the momentum transfers are small. We are therefore able to use the completeness relation $\Sigma_f \phi^*_f(\mathbf{r}) \phi_f(\mathbf{r}') = \delta(\mathbf{r}-\mathbf{r}')$ in summing over all final states in Eq. (9). In this way we find

$$(d\sigma/d\Omega)_2 = \langle |F(\mathbf{q},k,\mathbf{r})|^2 \rangle \tag{10}$$

which may be expressed in terms of the total phase shift by means of Eq. (2) and takes the form

$$\left(\frac{d\sigma}{d\Omega}\right)_2 = \left(\frac{k}{2\pi}\right)^2 \int |\phi_i(\mathbf{r})|^2 e^{i\mathbf{q}\cdot(\mathbf{b}-\mathbf{b}')} [1 - e^{i\chi_{tot}(\mathbf{b},\mathbf{r})}][1 - e^{-i\chi^*_{tot}(\mathbf{b}',\mathbf{r})}] d^{(2)}\mathbf{b} \, d^{(2)}\mathbf{b}' \, d\mathbf{r}. \tag{11}$$

This differential cross section, we should emphasize, furnishes the distribution of directions taken by the final center of mass momentum of the two emergent nucleons. It does not in general represent the angular distributions of the emerging nucleons considered individually.

The intensity of the diffraction-induced dissociation of the deuteron into a free neutron and a free proton, $(d\sigma/d\Omega)_{dis}$, is simply the difference between the summed scattered intensity for the emergence of two nucleons and the intensity of elastic scattering.

$$(d\sigma/d\Omega)_{dis} = (d\sigma/d\Omega)_2 - (d\sigma/d\Omega)_{el} \tag{12}$$

$$= \langle |F(\mathbf{q},k,\mathbf{r})|^2 \rangle - |\langle F(\mathbf{q},k,\mathbf{r}) \rangle|^2. \tag{13}$$

If we use Eqs. (8), (11), and (12) we may write this intensity in the form

$$\left(\frac{d\sigma}{d\Omega}\right)_{dis} = \left(\frac{k}{2\pi}\right)^2 \int e^{i\mathbf{q}\cdot(\mathbf{b}-\mathbf{b}')}\{\langle [1-e^{i\chi_{tot}(\mathbf{b},\mathbf{r})}][1-e^{-i\chi_{tot}^*(\mathbf{b}',\mathbf{r})}] \rangle$$
$$-\langle 1-e^{i\chi_{tot}(\mathbf{b},\mathbf{r})} \rangle\langle 1-e^{-i\chi_{tot}^*(\mathbf{b}',\mathbf{r})} \rangle\} d^{(2)}\mathbf{b}\, d^{(2)}\mathbf{b}' \tag{14}$$

$$= \left(\frac{k}{2\pi}\right)^2 \int e^{i\mathbf{q}\cdot(\mathbf{b}-\mathbf{b}')}\{\langle e^{i[\chi_{tot}(\mathbf{b},\mathbf{r})-\chi_{tot}^*(\mathbf{b}',\mathbf{r})]} \rangle - \langle e^{i\chi_{tot}(\mathbf{b},\mathbf{r})} \rangle\langle e^{-i\chi_{tot}^*(\mathbf{b}',\mathbf{r})} \rangle\} d^{(2)}\mathbf{b}\, d^{(2)}\mathbf{b}'. \tag{15}$$

The various intensities we have described may be integrated over all the directions of $\mathbf{k}'$, where $\hbar\mathbf{k}'$ is the momentum of the center of mass of the scattered two-particle system. The cross section for elastic scattering, $\sigma_{el}$, is given by

$$\sigma_{el} = \int \left(\frac{d\sigma}{d\Omega}\right)_{el} d\Omega_{\mathbf{k}'}. \tag{16}$$

Since high-energy scattering is predominantly small-angle scattering, the integration over the surface of the sphere $|\mathbf{k}'|^2 = k$ may be approximated by an integration over the plane, in momentum space, which is tangent to the sphere at the forward direction $\mathbf{k}'=\mathbf{k}$. The solid angle $d\Omega_{\mathbf{k}'}$ may therefore be represented approximately by $d^{(2)}\mathbf{k}'/k^2$, where $d^{(2)}\mathbf{k}'$ lies in the tangent plane mentioned. Since the angular distribution is a function of $\mathbf{q}=\mathbf{k}-\mathbf{k}'$ and $\mathbf{k}$ is fixed we may equivalently replace $d\Omega_{\mathbf{k}'}$ by $d^{(2)}\mathbf{q}/k^2$. With this approximation we carry out the integration in Eq. (16) and write the elastic scattering cross section as

$$\sigma_{el} = \int \left(\frac{d\sigma}{d\Omega}\right)_{el} \frac{d^{(2)}\mathbf{q}}{k^2}. \tag{17}$$

This integrated cross section may be expressed as an impact parameter integral in terms of the total phase shift $\chi_{tot}(\mathbf{b}, \mathbf{r})$ by means of Eq. (8) and by making use of the Fourier integral representation of the two-dimensional delta function. The result may be written as

$$\sigma_{el} = \int |\langle 1 - e^{i\chi_{tot}(\mathbf{b},\mathbf{r})} \rangle|^2 d^{(2)}\mathbf{b}. \tag{18}$$

The integrated cross section for the emergence of two nucleons, $\sigma_2$, is similarly given approximately by the integrals

$$\sigma_2 = \int \left(\frac{d\sigma}{d\Omega}\right)_2 \frac{d^{(2)}\mathbf{q}}{k^2} \tag{19}$$

$$= \int \langle |1 - e^{i\chi_{tot}(\mathbf{b},\mathbf{r})}|^2 \rangle d^{(2)}\mathbf{b}. \tag{20}$$

The integrated cross section for diffraction induced dissociation $\sigma_{dis}$ is easily obtained by integrating the corresponding intensity given by Eq. (14),

$$\sigma_{dis} = \int \left(\frac{d\sigma}{d\Omega}\right)_{dis} \frac{d^{(2)}\mathbf{q}}{k^2} \tag{21}$$

$$= \int [\langle |1 - e^{i\chi_{tot}(\mathbf{b},\mathbf{r})}|^2 \rangle - |\langle 1 - e^{i\chi_{tot}(\mathbf{b},\mathbf{r})} \rangle|^2 ] d^{(2)}\mathbf{b} \tag{22}$$

$$= \int [\langle |e^{i\chi_{tot}(\mathbf{b},\mathbf{r})}|^2 \rangle - |\langle e^{i\chi_{tot}(\mathbf{b},\mathbf{r})} \rangle|^2 ] d^{(2)}\mathbf{b}. \tag{23}$$

The optical theorem states that the deuteron total cross section, $\sigma_d$, is given by

$$\sigma_d = (4\pi/k) \, \text{Im} \, [F_{ii}(0, k)]. \tag{24}$$

When Eqs. (2) and (3) are used to evaluate the forward scattering amplitude in this expression we find

$$\sigma_d = 2\int [1 - \text{Re}\langle e^{i\chi_{tot}(\mathbf{b},\mathbf{r})} \rangle] d^{(2)}\mathbf{b}. \tag{25}$$

A simple check on the unitary character of these approximate expressions may be made by assuming that the target nucleus is represented by a real optical potential or a real phase shift function $\chi_{tot}(\mathbf{b}, \mathbf{r})$. In that case no excitations of the nucleus which are represented in the optical model by absorptive processes would occur. It is immediately seen that when $\chi_{tot}$ is real the two integrated cross sections $\sigma_{el}$ for elastic scattering and $\sigma_{dis}$ for diffraction dissociation sum to $\sigma_d$ as given by Eq. (25).

### 3. Semi-classical description of diffraction dissociation

Diffraction dissociation and other stripping processes can be understood, given some approximations, on a semi-quantitative basis in semi-classical terms. In this section we shall present such a description of the mechanism which induces the diffraction dissociation of the deuteron. This will be a discussion of the extreme case in which the size of the deuteron is much smaller than the dimension within which the nuclear field to which the deuteron is subject changes appreciably. One of the advantages of this discussion is that it allows us to make contact with simple classical images. The field of the nucleus does not, of course, quantitatively obey our restrictions. The surface thickness is generally smaller than the average deuteron radius. These approximate considerations only furnish a crude guide, but nonetheless are interesting because, for example, they show where in the nuclear potential the diffraction dissociation tends to predominantly take place.

For simplicity we shall assume that the effective total phase shift function $\chi_{tot}(\mathbf{b}, \mathbf{r})$ for deuteron-nucleus collisions may be approximated by the sum of the neutron-nucleus and proton-nucleus effective phase shifts $\chi_n$ and $\chi_p$ so that

$$\chi_{tot}(\mathbf{b}, \mathbf{r}) = \chi_n(\mathbf{b} - \tfrac{1}{2}\mathbf{s}) + \chi_p(\mathbf{b} + \tfrac{1}{2}\mathbf{s}), \tag{26}$$

where $\mathbf{s}$ is the projection of $\mathbf{r}$ on the plane of impact parameters, i.e., perpendicular to the direction of the incident beam. For neutron-nucleus and proton-nucleus effective forces with azimuthal symmetry about the direction of the incident beam, we may write $\chi_j(\mathbf{b}) = \chi_j(b)$ for j=n, p. If $\chi_j(b)$ tends to vary slowly over a distance equal to the average deuteron radius, as for example might be the case if the nucleus were very large, we might expand $\chi_j(\mathbf{b} \pm \frac{1}{2}\mathbf{s})$, for $j$=n, p, in a Taylor series as

$$\chi_j(\mathbf{b} \pm \tfrac{1}{2}\mathbf{s}) = \chi_j(b) \pm \tfrac{s}{2}\hat{\mathbf{s}} \cdot \hat{\mathbf{b}} \frac{\partial \chi_j(b)}{\partial b} + \tfrac{s^2}{8}\{(\hat{\mathbf{s}} \cdot \hat{\mathbf{b}})^2 \left[\frac{\partial^2 \chi_j(b)}{\partial b^2} - \frac{1}{b}\frac{\partial \chi_j(b)}{\partial b}\right] + \frac{1}{b}\frac{\partial \chi_j(b)}{\partial b}\} + \cdots, \qquad (27)$$

where $\hat{\mathbf{s}}$ and $\hat{\mathbf{b}}$ are unit vectors. The total phase shift, which is given by Eq. (26), may then be written as

$$\chi_n(\mathbf{b} - \tfrac{1}{2}\mathbf{s}) + \chi_p(\mathbf{b} + \tfrac{1}{2}\mathbf{s}) = \chi_n(b) + \chi_p(b) + \tfrac{s}{2}\hat{\mathbf{s}} \cdot \hat{\mathbf{b}}\frac{\partial}{\partial b}[\chi_p(b) - \chi_n(b)] +$$

$$+ \tfrac{s^2}{8}\{(\hat{\mathbf{s}} \cdot \hat{\mathbf{b}})^2\{\frac{\partial^2}{\partial b^2}[\chi_n(b) + \chi_p(b)] - \frac{1}{b}\frac{\partial}{\partial b}[\chi_n(b) + \chi_p(b)]\} + \frac{1}{b}\frac{\partial}{\partial b}[\chi_n(b) + \chi_p(b)]\} + \cdots. \qquad (28)$$

At high energies the effective phase shifts $\chi_n$ and $\chi_p$ may be generally described in terms of effective complex potentials $V_n$ and $V_p$ by [1]

$$\chi_j(b) = -\frac{1}{\hbar v}\int_{-\infty}^{\infty} V_j(b,z) dz, \qquad j\text{=n, p}, \qquad (29)$$

where **z** is the projection vector of the incident nucleon $j$ along the direction of the incident beam, i.e., parallel to **k**. The first derivative of $\chi_j(b)$ may therefore be expressed as

$$\chi_j'(b) = -\frac{1}{\hbar v}\int_{-\infty}^{\infty} \frac{\partial V_j(b,z)}{\partial b} dz. \qquad (30)$$

*Classically*, for the case of real potentials we may think of the gradient $-\partial V_j/\partial b$ as a radial force $(F_j)_b$ in the direction perpendicular to the incident beam. Then if we write $dz = v\,dt$, we may rewrite Eq. 30) as

$$\chi_j'(b) = \tfrac{1}{\hbar}\int_{-\infty}^{\infty} (F_j)_b dt \qquad (31)$$

$$= \tfrac{1}{\hbar} J_j(b). \qquad (32)$$

It is clear from this form that $\chi_j'$ corresponds to the radial impulse $J_j$ which is imparted to nucleon $j$ in its passage through the potential $V_j$. Similarly we have

$$\chi_j''(b) = \tfrac{1}{\hbar}\frac{\partial}{\partial b}\int_{-\infty}^{\infty} (F_j)_b dt \qquad (33)$$

$$= \tfrac{1}{\hbar}\frac{\partial}{\partial b}[J_j(b)] \qquad (34)$$

which represents the derivative of the radial impulse. Eq. (28) may therefore be written as

$$\chi_n(\mathbf{b} - \tfrac{1}{2}\mathbf{s}) + \chi_p(\mathbf{b} + \tfrac{1}{2}\mathbf{s}) = \chi_n(b) + \chi_p(b) + \tfrac{s}{2\hbar}\hat{\mathbf{s}} \cdot \hat{\mathbf{b}}[J_p(b) - J_n(b)] +$$

$$+ \tfrac{s^2}{8\hbar}\{(\hat{\mathbf{s}} \cdot \hat{\mathbf{b}})^2\{\frac{\partial}{\partial b}[J_n(b) + J_p(b)] - \frac{1}{b}[J_n(b) + J_p(b)]\} + \frac{1}{b}[J_n(b) + J_p(b)]\} + \cdots. \qquad (35)$$

The possibility of change in the internal state of the deuteron during the collision process arises from the dependence of this total phase shift on the coordinate **s**. If the **s**-dependent terms of the expansion are negligible in magnitude, no dissociation takes place through the diffraction mechanism.

The diffraction dissociation process is governed by $\langle |e^{i\chi_{tot}(b,r)}|^2 \rangle - |\langle e^{i\chi_{tot}(b,r)} \rangle|^2$, the integrand of Eq. (23). The first contribution to diffraction dissociation arises from the term linear in $s$ in Eq. (35) and depends on the *difference* between the two radial impulses, $J_p(b)$ and $J_n(b)$, which are communicated to the proton and neutron, respectively. If, for example, the neutron-nucleus and proton-nucleus interactions are assumed to be identical, $J_p - J_n$ vanishes and the term proportional to $s^2$ in Eq. (35) becomes the dominant s-dependent term. In that case Eq. (35) may be approximated by

$$\chi_n(\mathbf{b} - \tfrac{1}{2}\mathbf{s}) + \chi_p(\mathbf{b} + \tfrac{1}{2}\mathbf{s}) \approx \chi_n(b) + \chi_p(b) + \frac{s^2}{8\hbar}\{(\hat{\mathbf{s}} \cdot \hat{\mathbf{b}})^2 \{\frac{\partial}{\partial b}[J_n(b) + J_p(b)] +$$

$$- \frac{1}{b}[J_n(b) + J_p(b)]\} + \frac{1}{b}[J_n(b) + J_p(b)]\}. \quad (36)$$

We see that it is the total radial impulse, $J_n(b) + J_p(b)$, and its *derivative*, $\partial[J_n(b) + J_p(b)]/\partial b$, which govern the dissociation process at small scattering angles of the center of mass.

Since for nuclei which are not too heavy for which the neutron and proton forces are reasonably similar, it may be of some interest to pursue this approximation further and see thereby in what part of the nucleus the deuteron break-up takes place. If, for simplicity, we assume $V_n = V_p = V$, we may write $\chi_n = \chi_p = \chi$ and

$$\chi_n(\mathbf{b} - \tfrac{1}{2}\mathbf{s}) + \chi_p(\mathbf{b} + \tfrac{1}{2}\mathbf{s}) \approx 2\chi(b) + \frac{1}{4\hbar}\{(\mathbf{s} \cdot \hat{\mathbf{b}})^2[J'(b) - \tfrac{1}{b}J(b)] + s^2 \tfrac{1}{b}J(b)\}. \quad (37)$$

With this approximation we may write

$$b[\langle |e^{i\chi_{tot}(b,r)}|^2 \rangle - |\langle e^{i\chi_{tot}(b,r)} \rangle|^2] \propto \{[\tfrac{9}{5}\langle r^4 \rangle - \langle r^2 \rangle^2]b[J'(b) + \tfrac{1}{b}J(b)]^2 - \tfrac{12}{5}\langle r^4 \rangle J(b)J'(b)\}. \quad (38)$$

We note that the approximation we have used is that the average deuteron radius is much smaller than the distance in which $\chi$ varies appreciably. It is not a good approximation but it is an interesting guide as an idealization.

Let us consider the effective potential to be the Saxon-Woods potential

$$V(r) = -V_0/\{1 + \exp[(r - R_N)/a]\}. \quad (39)$$

The function $\chi'(b)$ may be written in the form

$$\chi'(b) = -\frac{2}{\hbar v}\int_0^\infty \frac{b}{r}V'(r)dz, \quad (40)$$

where $r = (b^2 + z^2)^{1/2}$. Similarly, the function $\chi''(b)$ may be written in the form

$$\chi''(b) = -\frac{2}{\hbar v}\int_0^\infty \left[\frac{b^2}{r^2}V''(r) + \frac{z^2}{r^3}V'(r)\right]dz. \quad (41)$$

With the Saxon-Woods potential, Eq. (40) becomes

$$\chi'(b) = -\frac{V_0}{\hbar v a}\int_0^\infty \frac{b}{r}\frac{1}{1+\cosh[(r-R_N)/a]}dz. \quad (42)$$

Similarly, Eq. (41) becomes

$$\chi''(b) = \frac{V_0}{\hbar v a} \int_0^\infty \left[ \frac{b^2}{ar^2} \frac{\sinh[(r-R_N)/a]}{\{1+\cosh[(r-R_N)/a]\}^2} - \frac{z^2}{r^3} \frac{1}{\{1+\cosh[(r-R_N)/a]\}} \right] dz. \qquad (43)$$

We define the right hand side of (38) as $g(b)$ which we may rewrite as

$$g(b) = C_1 b(\chi'' + \chi'/b)^2 - C_2 \chi' \chi'', \qquad (44)$$

with $C_1 = 9\langle r^4 \rangle/5 - \langle r^2 \rangle^2$ and $C_2 = 12\langle r^4 \rangle/5$. To go further we need an estimate of $C_2/C_1$. We shall use the Hulthén wave function for the deuteron ground state (ignoring the D-state for simplicity),

$$\phi_i(\mathbf{r}) = N (e^{-\gamma r} - e^{-\beta r})/r \qquad (45)$$

with $\gamma = 0.2316$ and $\beta = 5.98\gamma = 1.385$. The normalization constant is $N^2 = \gamma\beta(\gamma+\beta)/[2\pi(\beta-\gamma)^2] = 0.06204$.

For this wave function $C_2/C_1 \approx 1.5566$. Therefore

$$g(b) = C_1[b(\chi'' + \chi'/b)^2 - 1.5566 \chi' \chi'']. \qquad (46)$$

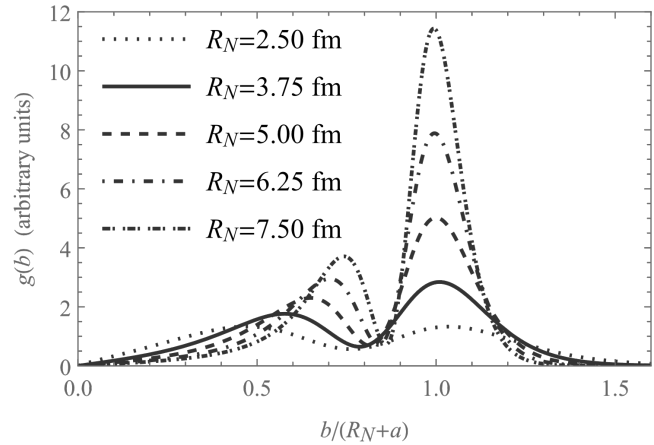

We have evaluated the right hand side of Eq. (46) numerically and in Figure 1 we present the results for the illustrative cases in which the parameter $R_N$ is taken to be 2.50, 3.75, 5.00, 6.25, and 7.50 fm. The skin thickness parameter, $a$, is taken to be 0.65 fm for all cases. We note that the absolute maximum occurs at $b/(R_N+a) \approx 1$ for each case except that for the lightest nucleus there is a second maximum of comparable magnitude at $b/(R_N+a) \approx 0.42$. At $b/(R_N+a) \approx 1$ we have $r \gtrsim R_N + a$ and $V(r)/V(0) \lesssim 0.27$, indicating that the diffraction dissociation process occurs predominantly in the surface region, outside the nuclear radius $R_N$. An alternative description would be to note, by integrating over $b$, that more than half the cross section comes from the region in which $b/(R_N+a)$ is greater than between 0.89 to 0.97 for all cases, except that for the lightest nucleus it is the region $b/(R_N+a) \gtrsim 0.74$ which contributes more than half the cross section. In the former cases $r$ is greater than between 0.89 to 0.97 times $(R_N + a)$ and $V(r)/V(0)$ is less than between 0.34 to 0.41, again indicating that the diffraction dissociation process occurs predominantly in the surface region, outside the nuclear radius $R_N$.

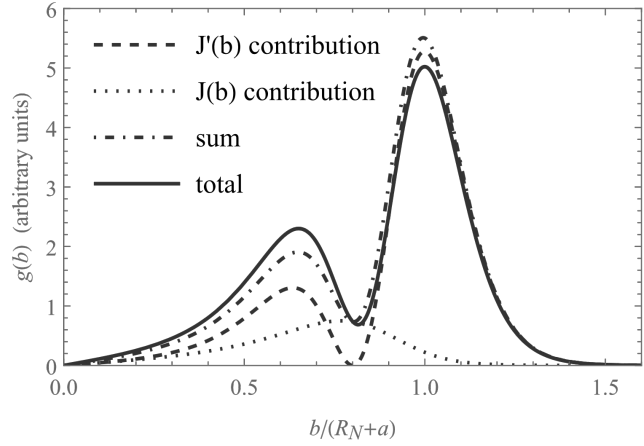

In Figure 2, for a medium-sized nucleus, $R_N$=5 fm, we show the contribution to $g(b)$ from only the radial impulse, $J(b)$, the contribution from only the radial derivative of the impulse, $J'(b)$, the sum of those two contributions, and the total magnitude of $g(b)$, which includes the interference terms involving both $J(b)$ and $J'(b)$. We see that the diffraction dissociation process is governed primarily by the derivative of the radial impulse. This result holds for all but the lightest nuclei and is more pronounced the heavier the nucleus.

## 4. Effects of absorption

Let us first consider the effect of inelastic collisions between a beam of single particles and nuclei. Inelastic collisions, in effect, remove particles from the beam and the change of energy means that they no longer interfere with the incident or elastically scattered waves. When the detection process is selective in energies in favor of elastic scattering, then these inelastically scattered particles are undetected and may as well have been described as having been absorbed. Such a description has been shown [1] to be possible quite generally where the procedure for finding the complex potential which is effective at high energies is described. The part of the complex potential which describes the absorption process is the imaginary part. By looking at the net flux of particles in the region where the imaginary part of the potential does not vanish, it is easy to show [1] that for a potential $V_1(\mathbf{r}_1)$ the absorption cross section, $\sigma_{1\text{abs}}$, is

$$\sigma_{1\text{abs}} = -\frac{2}{\hbar v}\int \psi^*(\mathbf{r}_1)\,\text{Im}[V_1(\mathbf{r}_1)]\,\psi(\mathbf{r}_1)\,d\mathbf{r}_1, \qquad (47)$$

where $\psi(\mathbf{r}_1)$ is the single-particle wave function. In the high-energy approximation [1] this wave function is

$$\psi(\mathbf{r}_1) = e^{i\mathbf{k}_1\cdot\mathbf{r}_1} e^{-\frac{i}{\hbar v}\int_{-\infty}^{z_1} V_1(\mathbf{b}_1+z'_1)dz'_1}, \qquad (48)$$

where we have written $\mathbf{r}_1=\mathbf{b}_1+\mathbf{z}_1$ with $\mathbf{z}_1$ being the projection of $\mathbf{r}_1$ along the direction of the incident beam, and where $\hbar\mathbf{k}_1$ is the momentum of the incident particle. If we substitute the wave function (48) into Eq. (47) and write the element of volume $d\mathbf{r}_1$ as $d^{(2)}\mathbf{b}_1 dz_1$ we obtain

$$\sigma_{1abs}= -\frac{2}{h v}\int d^{(2)}\mathbf{b}_1 \int_{-\infty}^{\infty} e^{\frac{2}{h v}\int_{-\infty}^{z_1} \text{Im}[V_1(\mathbf{b}_1+\mathbf{z}_1')]dz_1'} \text{Im}[V_1(\mathbf{b}_1+\mathbf{z}_1)]dz_1. \tag{49}$$

The $z_1$-integration is simply that of an exact differential and leads to

$$\sigma_{1abs}= \int [1- e^{\frac{2}{h v}\text{Im}\int_{-\infty}^{\infty} V_1(\mathbf{b}+\mathbf{z})dz}]d^{(2)}\mathbf{b} \tag{50}$$

$$= \int [1-|e^{i\chi_1(\mathbf{b})}|^2]d^{(2)}\mathbf{b}, \tag{51}$$

where $\chi_1(\mathbf{b})$ is given by Eq. (29).

A deuteron-nucleus absorption process is a process in which some incoherent collision takes place involving either the neutron or the proton or both. The concept of absorption for the deuteron-nucleus collision is a generalization of the idea of absorption for the single-particle-nucleus problem. To discuss absorption processes for the deuteron we shall assume that the nucleus may be represented by an optical model as far as each of the nucleon-nucleus interactions is concerned. That is to say, the neutron and the proton are each assumed to interact with suitably determined but independent optical model representations of the nucleus. We let the optical potential seen by the neutron be $V_n(\mathbf{r}_n)$ and that seen by the proton $V_p(\mathbf{r}_p)$. We shall approximate the total optical potential by adding the two independent interactions to say

$$V(\mathbf{r}_n, \mathbf{r}_p) = V_n(\mathbf{r}_n) + V_p(\mathbf{r}_p). \tag{52}$$

With this representation of the nuclear interactions we can find the absorption cross section $\sigma_{abs}$ for deuteron-nucleus collisions through a simple generalization of the formulae we have discussed for nucleon-nucleus collisions. This cross section, $\sigma_{abs}$, is one for collisions in which *one or both* of the particles is absorbed.

In place of a single-particle wave function we must now consider the two-particle wave function $\psi$ $(\mathbf{r}_n, \mathbf{r}_p)$, where $\mathbf{r}_n$ and $\mathbf{r}_p$ are position vectors for the neutron and proton in the deuteron. In the expression for the absorption cross section of the deuteron, $\sigma_{abs}$, corresponding to Eq. (47) for the single-particle absorption cross section, $\sigma_{1abs}$, we use not $\text{Im}[V_1(\mathbf{r}_1)]$ but rather $\text{Im}[V(\mathbf{r}_n, \mathbf{r}_p)]$. By means of Eq. (52) we may write the absorption cross section of the deuteron as

$$\sigma_{abs} = -\frac{2}{h v}\int \psi^*(\mathbf{r}_n, \mathbf{r}_p)\, \text{Im}[\,V_n(\mathbf{r}_n)+ V_p(\mathbf{r}_p)]\psi(\mathbf{r}_n, \mathbf{r}_p)d\mathbf{r}_n d\mathbf{r}_p. \tag{53}$$

It will be convenient to make the abbreviation

$$\omega_j(\mathbf{r}_j) = -\frac{1}{h v}\int_{-\infty}^{z_j} V_j(\mathbf{b}_j+\mathbf{z}_j')\, dz_j', \qquad j=n, p. \tag{54}$$

If we let $\mathbf{R} = \frac{1}{2}(\mathbf{r}_p+\mathbf{r}_n)$ be the position vector of the deuteron center of mass, the two-particle wave function is given in the high-energy approximation by

$$\psi(\mathbf{r}_n, \mathbf{r}_p) = e^{i\mathbf{k}\cdot\mathbf{R}+i[\omega_n(\mathbf{r}_n)+\omega_p(\mathbf{r}_p)]}\phi(\mathbf{r}_p - \mathbf{r}_n). \tag{55}$$

The squared modulus of this wave function is given by

$$|\psi(\mathbf{r}_n, \mathbf{r}_p)|^2 = e^{-2\text{Im}[\omega_n(\mathbf{r}_n)+\omega_p(\mathbf{r}_p)]}|\phi(\mathbf{r}_p - \mathbf{r}_n)|^2. \tag{56}$$

The position vectors $\mathbf{r}_p$ and $\mathbf{r}_n$ of the proton and neutron are related to the center-of-mass and relative coordinates $\mathbf{R}$ and $\mathbf{r}$ of the deuteron by

$$\mathbf{r}_p = \mathbf{R} + \tfrac{1}{2}\mathbf{r} \tag{57}$$

$$\mathbf{r}_n = \mathbf{R} - \tfrac{1}{2}\mathbf{r}. \tag{58}$$

If we substitute Eq. (56) into Eq. (53), transform the integration variables to $\mathbf{R}$ and $\mathbf{r}$, and replace $d\mathbf{R}$ by $d^{(2)}\mathbf{b}\,dz$, we obtain

$$\sigma_{\text{abs}} = -\frac{2}{\hbar v}\int|\phi(\mathbf{r})|^2 d\mathbf{r}\int \text{Im}[V_n(\mathbf{R}-\tfrac{1}{2}\mathbf{r})+V_p(\mathbf{R}+\tfrac{1}{2}\mathbf{r})]e^{-2\text{Im}[\omega_n(\mathbf{R}-\tfrac{1}{2}\mathbf{r})+\omega_p(\mathbf{R}+\tfrac{1}{2}\mathbf{r})]}dz\,d^{(2)}\mathbf{b}. \tag{59}$$

The z-integration is once again that of an exact differential and results in a simple expression for the deuteron absorption cross section given by

$$\sigma_{\text{abs}} = \int[1-|e^{i\chi_n(\mathbf{b}-\tfrac{1}{2}\mathbf{s})+i\chi_p(\mathbf{b}+\tfrac{1}{2}\mathbf{s})}|^2]|\phi(\mathbf{r})|^2 d\mathbf{r}\,d^{(2)}\mathbf{b}. \tag{60}$$

The interpretation of this result as a cross section for all absorption processes is confirmed by adding this expression to that given by Eq. (20) for the cross section $\sigma_2$ which is the cross section for the emergence of two particles, whether bound or unbound. We find that the sum is just the deuteron total cross section $\sigma_d$,

$$\sigma_d = \sigma_2 + \sigma_{\text{abs}}. \tag{61}$$

The absorption processes which may occur in deuteron-nucleus collisions are of three varieties. There are those in which the neutron is absorbed and not the proton, those in which the proton is absorbed and not the neutron, and finally those in which both the neutron and the proton are absorbed. We shall be interested in finding expressions for each of these cross sections considered individually. In order to do this it is convenient to begin by dividing the total absorption cross section $\sigma_{\text{abs}}$ into two parts according to whether or not a particular particle, say the proton, is absorbed. We shall let $\sigma_{\text{pabs}}$ denote the cross section corresponding to processes in which the proton is absorbed whether the neutron is absorbed or not. The remainder of the total absorption cross section, which we write as $\sigma_{\text{pas}}$, is the cross section for processes in which the neutron is absorbed and the *proton* remains free, i.e., Serber-type *absorptive stripping*.

Let us now derive the cross section $\sigma_{\text{pabs}}$ for absorption of the proton with no attention being paid to the neutron. This derivation will also furnish us with a useful way of writing the free-proton absorption cross section. In considering $\sigma_{\text{pabs}}$ we are interested only in what happens to the proton in collisions in which the deuteron is the projectile and we ignore the behavior of the neutron. Hence we are considering only a part of a two-particle system. Before any interaction between the deuteron and the nucleus takes place, the deuteron as a whole may be described by the two-particle incident wave function $\psi_{inc}(\mathbf{r}_n, \mathbf{r}_p)$ given by

$$\psi_{inc}(\mathbf{r}_n, \mathbf{r}_p) = e^{i\frac{1}{2}\mathbf{k}\cdot(\mathbf{r}_n+\mathbf{r}_p)}\phi_i(\mathbf{r}_p - \mathbf{r}_n). \tag{62}$$

This wave function cannot generally be written as a product of a function of $\mathbf{r}_p$ alone and a function of $\mathbf{r}_n$ alone; so the proton does not have its own wave function. The use of this pure state wave function is equivalent to the use of a two-particle density matrix

$$\psi_{inc}^*(\mathbf{r}_n', \mathbf{r}_p')\psi_{inc}(\mathbf{r}_n, \mathbf{r}_p). \tag{63}$$

Discussions which deal *only* with the way in which the *proton* is affected by the collision and which ignore the behavior of the neutron can best be carried out by using a reduced form of the density matrix which depends only upon the proton coordinates. The reduced density operator for the proton corresponding to the incident deuteron wave is given by

$$\rho_{pinc}(\mathbf{r}_p', \mathbf{r}_p) = \int \psi_{inc}^*(\mathbf{r}_n, \mathbf{r}_p')\psi_{inc}(\mathbf{r}_n, \mathbf{r}_p)d\mathbf{r}_n \tag{64}$$

$$= e^{i\frac{1}{2}\mathbf{k}\cdot(\mathbf{r}_p-\mathbf{r}_p')}\int \phi_i^*(\mathbf{r}_p' - \mathbf{r}_n)\phi_i(\mathbf{r}_p - \mathbf{r}_n)d\mathbf{r}_n . \tag{65}$$

To treat by means of the high-energy approximation high-energy deuteron collisions in which the proton is absorbed, we require the reduced density matrix $\rho_p(\mathbf{r}_p', \mathbf{r}_p)$ in that approximation. We know from Eqs. (48) and (54) that the single-particle wave function may be written as

$$\psi_p(\mathbf{r}_p) = e^{i\mathbf{k}_p\cdot\mathbf{r}_p}e^{i\omega_p(\mathbf{r}_p)} . \tag{66}$$

Similarly, we may show that $\rho_p(\mathbf{r}_p', \mathbf{r}_p)$ may be written as

$$\rho_p(\mathbf{r}_p', \mathbf{r}_p) = e^{-i\omega_p^*(\mathbf{r}_p')}\rho_{pinc}(\mathbf{r}_p', \mathbf{r}_p)e^{i\omega_p(\mathbf{r}_p)}. \tag{67}$$

The absorption cross section for a particle with a known wave function we have seen is given by Eq. (47). To calculate $\sigma_{pabs}$ we note that for a particle described by the reduced density matrix $\rho_p(\mathbf{r}_p', \mathbf{r}_p)$ the wave function product $\psi^*(\mathbf{r}_p)\psi(\mathbf{r}_p)$ in Eq.(47) must more generally be replaced by the diagonal matrix element $\rho_p(\mathbf{r}_p, \mathbf{r}_p)$. Hence the absorption cross section for the proton becomes

$$\sigma_{pabs} = -\frac{2}{\hbar v}\int Im[V_p(\mathbf{r}_p)]\rho_p(\mathbf{r}_p, \mathbf{r}_p)d\mathbf{r}_p. \tag{68}$$

The diagonal element of the density matrix $\rho_p(\mathbf{r}_p, \mathbf{r}_p)$ is simply given by

$$\rho_p(\mathbf{r}_p, \mathbf{r}_p) = e^{-2Im[\omega_p(\mathbf{r}_p)]}\rho_{pinc}(\mathbf{r}_p, \mathbf{r}_p) \tag{69}$$

$$= e^{-2Im[\omega_p(\mathbf{r}_p)]}\int |\phi(\mathbf{r}_p - \mathbf{r}_n)|^2 d\mathbf{r}_n \tag{70}$$

$$= e^{-2Im[\omega_p(\mathbf{r}_p)]} . \tag{71}$$

Therefore the absorption cross section for the proton in deuteron-nucleus collisions is given by

$$\sigma_{\text{pabs}} = -\frac{2}{\hbar v} \int \text{Im}[V_p(\mathbf{r}_p)] e^{-2\text{Im}[\omega_p(\mathbf{r}_p)]} d\mathbf{r}_p \tag{72}$$

$$= \int [1 - |e^{i\chi_p(\mathbf{b}_p)}|^2] d^{(2)}\mathbf{b}_p. \tag{73}$$

We note that the last form for $\sigma_{\text{pabs}}$ is equal to the free-particle absorption cross section given in Eq. (51). This simple result should come as no surprise since by ignoring the fate of the neutron and eliminating its position coordinate from the calculation we are in effect simply dealing with the case in which a uniform flux of protons is incident upon the nucleus. Although the result in Eq. (73) is thus in a sense trivial, the means by which we have derived it is not and will shortly prove useful as we shall use an extension of this method in discussing the remaining part of the cross section ($\sigma_{\text{pas}}$).

It is useful to separate $\sigma_{\text{pabs}}$ itself into two parts. One cross section, $\sigma_{\text{2abs}}$, is that for two-particle absorption and represents processes in which both the proton and neutron are absorbed. The other cross section, $\sigma_{\text{nas}}$, represents stripping processes in which the proton is absorbed and the neutron emerges as a free particle. The reduced density matrix $\rho_p(\mathbf{r}_p', \mathbf{r}_p)$ which we introduced in Eq. (67) corresponds to a description of the proton *regardless* of what happens to the neutron. That is, the neutron *may or may not* be absorbed. The reduced density matrix $\rho_p'(\mathbf{r}_p', \mathbf{r}_p)$ which corresponds to situations in which the neutron *is not* absorbed is given by

$$\rho_p'(\mathbf{r}_p', \mathbf{r}_p) = \int \psi^*(\mathbf{r}_n, \mathbf{r}_p') \psi(\mathbf{r}_n, \mathbf{r}_p) d\mathbf{r}_n. \tag{74}$$

The difference

$$\bar{\rho}_p(\mathbf{r}_p', \mathbf{r}_p) = \rho_p(\mathbf{r}_p', \mathbf{r}_p) - \rho_p'(\mathbf{r}_p', \mathbf{r}_p) \tag{75}$$

then describes the proton when the neutron has been absorbed. Similarly

$$\bar{\rho}_n(\mathbf{r}_n', \mathbf{r}_n) = \rho_n(\mathbf{r}_n', \mathbf{r}_n) - \rho_n'(\mathbf{r}_n', \mathbf{r}_n) \tag{76}$$

describes the neutron when the proton has been absorbed. If a neutron is first absorbed in a collision, the rate at which the proton is absorbed is

$$-\frac{2}{\hbar} \int [\text{Im} V_p(\mathbf{r}_p)] \bar{\rho}_p(\mathbf{r}_p, \mathbf{r}_p) d\mathbf{r}_p. \tag{77}$$

Similarly, if as proton is first absorbed in a collision, the rate at which the neutron is absorbed is

$$-\frac{2}{\hbar} \int \text{Im}[V_n(\mathbf{r}_n)] \bar{\rho}_n(\mathbf{r}_n, \mathbf{r}_n) d\mathbf{r}_n. \tag{78}$$

The cross section $\sigma_{\text{2abs}}$ for two-particle absorption in which both the neutron and the proton are absorbed is obtained by adding the two rates so that

$$\sigma_{\text{2abs}} = -\frac{2}{\hbar v} \int \text{Im}[V_p(\mathbf{r}_p)] \bar{\rho}_p(\mathbf{r}_p, \mathbf{r}_p) d\mathbf{r}_p - \frac{2}{\hbar v} \int \text{Im}[V_n(\mathbf{r}_n)] \bar{\rho}_n(\mathbf{r}_n, \mathbf{r}_n) d\mathbf{r}_n. \tag{79}$$

By means of Eq. (70) we may write $\rho_p(\mathbf{r}_p, \mathbf{r}_p)$ as

$$\rho_p(\mathbf{r}_p, \mathbf{r}_p) = e^{i[\omega_p(\mathbf{r}_p) - \omega_p^*(\mathbf{r}_p)]} \int |\phi(\mathbf{r}_p - \mathbf{r}_n)|^2 d\mathbf{r}_n. \tag{80}$$

By means of Eqs. (55) and (74) we may write $\rho_p'(\mathbf{r}_p, \mathbf{r}_p)$ as

$$\rho_p'(\mathbf{r}_p, \mathbf{r}_p) = \int \psi^*(\mathbf{r}_n, \mathbf{r}_p)\,\psi(\mathbf{r}_n, \mathbf{r}_p)\,d\mathbf{r}_n \tag{81}$$

$$= e^{i[\omega_p(\mathbf{r}_p) - \omega_p^*(\mathbf{r}_p)]} \int e^{i[\omega_n(\mathbf{r}_n) - \omega_n^*(\mathbf{r}_n)]} |\phi(\mathbf{r}_p - \mathbf{r}_n)|^2 d\mathbf{r}_n. \tag{82}$$

The density matrix $\bar{\rho}_p(\mathbf{r}_p, \mathbf{r}_p)$ is therefore, by Eq. (75),

$$\bar{\rho}_p(\mathbf{r}_p, \mathbf{r}_p) = e^{i[\omega_p(\mathbf{r}_p) - \omega_p^*(\mathbf{r}_p)]} \int \{1 - e^{i[\omega_n(\mathbf{r}_n) - \omega_n^*(\mathbf{r}_n)]}\} |\phi(\mathbf{r}_p - \mathbf{r}_n)|^2 d\mathbf{r}_n \tag{83}$$

$$= |e^{i\omega_p(\mathbf{r}_p)}|^2 \int [1 - |e^{i\omega_n(\mathbf{r}_n)}|^2] |\phi(\mathbf{r}_p - \mathbf{r}_n)|^2 d\mathbf{r}_n. \tag{84}$$

The density matrix $\bar{\rho}_n(\mathbf{r}_n, \mathbf{r}_n)$ is given by Eq. (84) with the subscripts n and p interchanged. The two-particle absorption cross section may therefore be written as

$$\sigma_{2abs} = -\frac{2}{\hbar v} \int \mathrm{Im}[V_p(\mathbf{r}_p)]\,|e^{i\omega_p(\mathbf{r}_p)}|^2 d\mathbf{r}_p \int [1 - |e^{i\omega_n(\mathbf{r}_n)}|^2]|\phi(\mathbf{r}_p - \mathbf{r}_n)|^2 d\mathbf{r}_n$$

$$-\frac{2}{\hbar v} \int \mathrm{Im}[V_n(\mathbf{r}_n)]\,|e^{i\omega_n(\mathbf{r}_n)}|^2 d\mathbf{r}_n \int [1 - |e^{i\omega_p(\mathbf{r}_p)}|^2]|\phi(\mathbf{r}_p - \mathbf{r}_n)|^2 d\mathbf{r}_p \tag{85}$$

$$= -\frac{2}{\hbar v} \int |\phi(\mathbf{r}_p - \mathbf{r}_n)|^2 \{\mathrm{Im}[V_p(\mathbf{r}_p)] e^{-2\mathrm{Im}[\omega_p(\mathbf{r}_p)]} [1 - |e^{i\omega_n(\mathbf{r}_n)}|^2]$$

$$+ \mathrm{Im}[V_n(\mathbf{r}_n)] e^{-2\mathrm{Im}[\omega_n(\mathbf{r}_n)]} [1 - |e^{i\omega_p(\mathbf{r}_p)}|^2]\} d\mathbf{r}_p d\mathbf{r}_n. \tag{86}$$

We may rewrite the quantity inside the brackets { } in Eq.(86) as

$$\mathrm{Im}[V_p(\mathbf{r}_p)] e^{-2\mathrm{Im}[\omega_p(\mathbf{r}_p)]} + \mathrm{Im}[V_n(\mathbf{r}_n)] e^{-2\mathrm{Im}[\omega_n(\mathbf{r}_n)]} - \mathrm{Im}[V_p(\mathbf{r}_p) + V_n(\mathbf{r}_n)] e^{-2\mathrm{Im}[\omega_p(\mathbf{r}_p) + \omega_n(\mathbf{r}_n)]}.$$

If we transform the integration variables to the deuteron relative and center-of-mass coordinates $\mathbf{r}$ and $\mathbf{R}$ we obtain

$$\sigma_{2abs} = -\frac{2}{\hbar v} \int |\phi(\mathbf{r})|^2 \{\mathrm{Im}[V_p(\mathbf{R} + \tfrac{1}{2}\mathbf{r})] e^{-2\mathrm{Im}[\omega_p(\mathbf{R} + \tfrac{1}{2}\mathbf{r})]} + \mathrm{Im}[V_n(\mathbf{R} - \tfrac{1}{2}\mathbf{r})] e^{-2\mathrm{Im}[\omega_n(\mathbf{R} - \tfrac{1}{2}\mathbf{r})]}$$

$$- \mathrm{Im}[V_p(\mathbf{R} + \tfrac{1}{2}\mathbf{r}) + V_n(\mathbf{R} - \tfrac{1}{2}\mathbf{r})] e^{-2\mathrm{Im}[\omega_p(\mathbf{R} + \tfrac{1}{2}\mathbf{r}) + \omega_n(\mathbf{R} - \tfrac{1}{2}\mathbf{r})]}\} d\mathbf{r}\, d\mathbf{R} \tag{87}$$

$$= \int \langle 1 - |e^{i\chi_p(\mathbf{b} + \tfrac{1}{2}\mathbf{s})}|^2 + 1 - |e^{i\chi_n(\mathbf{b} - \tfrac{1}{2}\mathbf{s})}|^2 - \{1 - |e^{i[\chi_p(\mathbf{b} + \tfrac{1}{2}\mathbf{s}) + \chi_n(\mathbf{b} - \tfrac{1}{2}\mathbf{s})]}|^2\}\rangle d^{(2)}\mathbf{b} \tag{88}$$

$$= \int \langle [1 - |e^{i\chi_p(\mathbf{b} + \tfrac{1}{2}\mathbf{s})}|^2][1 - |e^{i\chi_n(\mathbf{b} - \tfrac{1}{2}\mathbf{s})}|^2]\rangle d^{(2)}\mathbf{b}. \tag{89}$$

We have now obtained the cross section $\sigma_{2abs}$ for absorption processes in which some incoherent collision takes place involving both the neutron and the proton. We note that this expression has the expected form for two-particle absorption, namely it contains a factor $1 - |e^{i\chi_p}|^2$ for absorption of the proton and a factor $1 - |e^{i\chi_n}|^2$ for absorption of the neutron.

As we have noted earlier, we could look only at those processes in which the proton is absorbed and regard the complete absorption cross section for the proton ($\sigma_{pabs}$) as the sum of two terms, one of which ($\sigma_{2abs}$) corresponds to both the proton and neutron being absorbed and the second ($\sigma_{nas}$) to only the proton being absorbed. It follows that the second of these terms describes the absorption of the proton with the neutron going free and is therefore a cross section for Serber-type stripping in which a neutron is liberated. Since we may write

$$\sigma_{pabs} = \sigma_{2abs} + \sigma_{nas} \qquad (90)$$

we obtain by means of Eqs. (73), (89), and (90) the relation

$$\sigma_{nas} = \int \langle \left| e^{i\chi_n(\mathbf{b}-\frac{1}{2}\mathbf{s})} \right|^2 [1 - \left| e^{i\chi_p(\mathbf{b}+\frac{1}{2}\mathbf{s})} \right|^2] \rangle d^{(2)}\mathbf{b} . \qquad (91)$$

We note that if the proton-nucleus optical potential is real, $\sigma_{nas}$ vanishes as expected. In a similar manner we find that the cross section $\sigma_{pas}$ for collisions in which the neutron is absorbed but the proton goes free is given by

$$\sigma_{pas} = \int \langle \left| e^{i\chi_p(\mathbf{b}+\frac{1}{2}\mathbf{s})} \right|^2 [1 - \left| e^{i\chi_n(\mathbf{b}-\frac{1}{2}\mathbf{s})} \right|^2] \rangle d^{(2)}\mathbf{b} . \qquad (92)$$

The total stripping cross section for emerging protons ($\sigma_{ps}$) is given by the sum of the cross section $\sigma_{dis}$ for diffraction dissociation and the cross section $\sigma_{pas}$ for absorptive stripping and may be written by means of Eqs. (23) and (92) as

$$\sigma_{ps} = \sigma_{dis} + \sigma_{pas} \qquad (93)$$

$$= \int [\langle \left| e^{i\chi_p(\mathbf{b}+\frac{1}{2}\mathbf{s})} \right|^2 \rangle - \left| \langle e^{i\chi_p(\mathbf{b}+\frac{1}{2}\mathbf{s}) + i\chi_n(\mathbf{b}-\frac{1}{2}\mathbf{s})} \rangle \right|^2 ] d^{(2)}\mathbf{b} . \qquad (94)$$

Similarly the total stripping cross section for emerging neutrons ($\sigma_{ns}$) is given by

$$\sigma_{ns} = \sigma_{dis} + \sigma_{nas} \qquad (95)$$

$$= \int [\langle \left| e^{i\chi_n(\mathbf{b}-\frac{1}{2}\mathbf{s})} \right|^2 \rangle - \left| \langle e^{i\chi_p(\mathbf{b}+\frac{1}{2}\mathbf{s}) + i\chi_n(\mathbf{b}-\frac{1}{2}\mathbf{s})} \rangle \right|^2 ] d^{(2)}\mathbf{b} . \qquad (96)$$

Finally, the cross section $\sigma_f$ for the breakup of the deuteron and the resulting emergence of one or two free nucleons is given by

$$\sigma_f = \sigma_{dis} + \sigma_{pas} + \sigma_{nas} \qquad (97)$$

$$= \int \left[ \langle \left| e^{i\chi_n(\mathbf{b}-\frac{1}{2}\mathbf{s})} \right|^2 + \left| e^{i\chi_p(\mathbf{b}+\frac{1}{2}\mathbf{s})} \right|^2 - \left| e^{i\chi_p(\mathbf{b}+\frac{1}{2}\mathbf{s}) + i\chi_n(\mathbf{b}-\frac{1}{2}\mathbf{s})} \right|^2 \rangle - \left| \langle e^{i\chi_p(\mathbf{b}+\frac{1}{2}\mathbf{s}) + i\chi_n(\mathbf{b}-\frac{1}{2}\mathbf{s})} \rangle \right|^2 \right] d^{(2)}\mathbf{b}. \qquad (98)$$

---

**Declaration of competing interest**

The living author has no known competing financial interests or personal relationship that could have appeared to influence the work reported in this paper.

**Acknowledgements**

This work was supported in part by the National Science Foundation, the National Aeronautics and Space Administration, and the City University of New York Faculty Research Award Program (V.F.), and by the Air Force Office of Scientific Research (R.J.G.).

**References**

[1] R. J. Glauber, in *Lectures in Theoretical Physics*, edited by W. E. Brittin *et al*. (IntersciencePublishers, Inc., New York, 1959), Vol. I, p. 315.

[2] V. Franco, Ph.D. thesis, Harvard University (1963) (unpublished).

[3] V. Franco and R. J. Glauber, Phys. Rev. 142, 1195 (1966). See also R. J. Glauber, Phys. Rev. 100, 242 (1955).

[4] L. Zamick, Ann. Phys. (N.Y.) 21, 550 (1963).

[5] G. Fäldt, Phys. Rev. D 2, 846 (1970).

[6] R. J. Glauber, Phys. Rev. 99, 1515 (1955).

[7] See Refs. [1]-[5] as well as, for example, D. R. Harrington, Phys. Rev. **135**, B358 (1964); **137**, AB3(E) (1965); **184**, 1745 (1969), V. Franco, Phys. Rev. Letters **16**, 944 (1966), W. Czyż and L. Leśniak, Phys. Letters **24B**, 227 (1967); W. Czyż and L.C. Maximon, Ann. Phys. (N.Y.) **52**, 59 (1969), V. Franco. Phys. Rev 140, B1501 (1965), Yu. L. Parfenova and M. V. Zhukov, Eur. Phys. J. **A12**, 191 (2001); A. Mehndiratta and P. Shukla, Nuclear Physics **A961**, 22 (2017).

[8] Application to atomic physics was first made in V. Franco, Phys. Rev. Letters **20**, (1968). Application to molecular physics was made, for example, in F. A. Gianturco and U. T. Lamanna, Molec. Phys. **40**, 793 (1980). Application to the quark model in particle physics was first made in V. Franco, Phys. Rev. Letters **18**, 1159 (1967).

**Figure Captions**

FIG. 1. The function $g(b)$, Eq. (46), for a Saxon-Woods effective potential, Eq. (39), as a function of $b/(R_N+a)$ for several "radii" $R_N$: 2.50 fm dotted line, 3.75 fm continuous line, 5.00 fm dashed line, 6.25 fm

dot-dashed line, 7.50 fm dot-dot-dashed line. The skin thickness parameter, $a$, was kept fixed at 0.65 fm. The absolute maxima occur very close to $b=R_N+a$.

FIG. 2. Contributions to $g(b)$, Eq. (46), from the radial impulse, $J(b)$, (dotted), the derivative of the radial impulse, $J'(b)$, (dashed), the sum of the two, (dot-dashed), and the total magnitude, (solid) which includes interference terms.

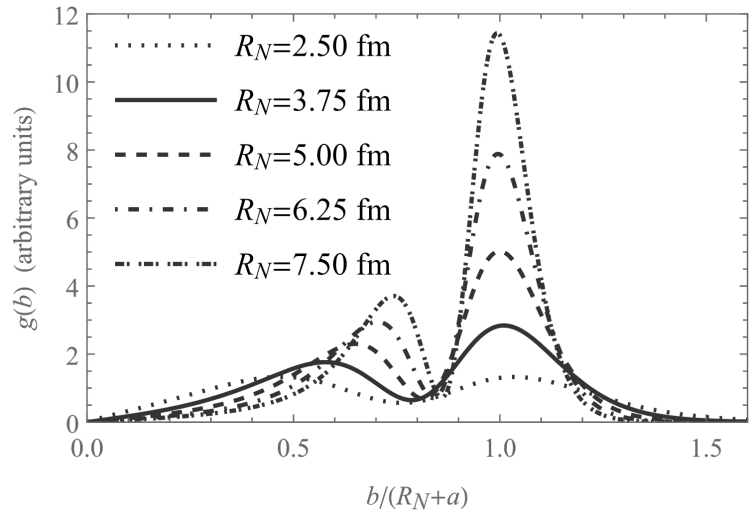
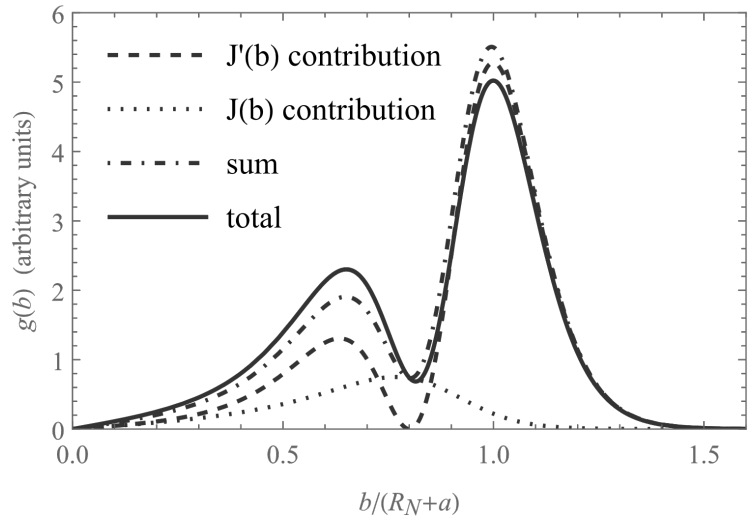